\documentclass[referee, useAMS, usenatbib]{mn2e}
\usepackage{psfig}
\usepackage{graphicx}
\usepackage{epsf}
\usepackage{bm}
\usepackage{lscape}

\title [Bridging between optical and infrared]
{Transformations between WISE, 2MASS, SDSS and BVRI photometric systems:
I. Transformation equations for dwarfs}

\author[Bilir et al.]
       {S. Bilir,${^1} \thanks{E-mail: sbilir@istanbul.edu.tr}$
        S. Karaali$^{2}$, S. Ak$^{1}$, N. D. Da\u gtekin$^{1}$, \"O. \"Onal$^{1}$, E. Yaz$^{1}$, \newauthor
        B. Co\c skuno\u glu$^{1}$, A. Cabrera-Lavers$^{3, 4}$          
\\
  $^1$Istanbul University Science Faculty, Department of Astronomy and Space 
Sciences, 34119, University-Istanbul, Turkey\\
  $^2$Beykent University, Faculty of Science and Letters, Department of Mathematics  
and Computer, Beykent 34398, Istanbul, Turkey\\
  $^3$Instituto de Astrof\'{\i}sica de Canarias, E-38205 La Laguna, Tenerife, Spain\\
  $^4$GTC Project Office, E-38205 La Laguna, Tenerife, Spain\\}

\pagerange{\pageref{firstpage}--\pageref{lastpage}} \pubyear{2007}

\begin{document}
\maketitle
\label{firstpage}
\begin{abstract}
We present colour transformations for the conversion of the $W1$ and $W2$ magnitudes 
of WISE photometric system to the Johnson – Cousins' {\it BVRI}, SDSS {\it gri}, and 2MASS
{\it JHK$_{s}$} photometric systems, for dwarfs. The $W3$ and $W4$ magnitudes were not 
considered due to their insufficient signal to noise ratio (S/N). The coordinates of 825 dwarfs 
along with their {\it BVRI}, {\it gri}, and {\it JHK$_{s}$} data, taken from \citet{Bilir08} 
were matched with the coordinates of stars in the preliminary data release of WISE 
\citep{Wright10} and a homogeneous dwarf sample with high S/N ratio have 
been obtained using the following constraints: 1) the data were dereddened, 2) giants 
were identified and excluded from the sample, 3) sample stars were selected according  
to data quality, 4) transformations were derived for sub samples of 
different metallicity range, and 5) transformations are two colour dependent. 
These colour transformations, coupled with known absolute magnitudes at shorter wavelenghts, can be 
used in space density evaluation for the Galactic (thin and thick) discs, at distances 
larger than the ones evaluated with {\it JHK$_{s}$} photometry.  
\end{abstract}

\begin{keywords}
surveys--catalogues--techniques: photometric
\end{keywords}

\section{Introduction}
All sky surveys from X-ray to radio regions of the electromagnetic spectrum have great impact on our understanding of Galactic structure and also change our view of the Universe dramatically. While surveys in optical wavelengths give detailed information about Galactic halo, longer wavelength surveys give us useful results for the Galactic disc. Recent photometric surveys such as Sloan Digital Sky Survey \citep[SDSS;][]{York00}, Two Micron All Sky Survey \citep[2MASS;][]{Skrutskie06}, and Wide field Infrared Survey Explorer \citep[WISE;][]{Wright10} are used to classify objects in colour spaces, while spectroscopic surveys such as SDSS and RAdial Velocity Experiment \citep[RAVE,][]{Steinmetz06} are used to determine stellar atmospheric parameters. Analysis of astrometric, radial velocity and stellar atmospheric data, especially in the Solar neighbourhood, let us understand the structure, formation, and evolution of the Galaxy. Deep sky surveys are powerful tools in examining the far side of our Galaxy. However, information on nearby space can be obtained from shallow wavelength surveys. Therefore, for a complete understanding of the whole Galaxy multiple systems are needed. This implies the use of transformation equations between various photometric systems. 

SDSS obtains images almost simultaneously in five broad bands ($u$, $g$, $r$, $i$, and $z$) centered at 3540, 4760, 6280, 7690, and 9250 \AA, respectively, \citep{Fukugita96, Gunn98, Hogg01, Smith02}. The magnitudes derived from fitting a point spread function (PSF) are currently accurate to about 2 per cent in $g$, $r$, and $i$, and 3-5 per cent in $u$ and $z$ for bright ($<$ 20 mag) point sources. The data have been made public in a series of yearly data release where the eighth \citep[DR8,][]{Aihara11} covers 14 555 deg$^{2}$ of imaging area. The limiting magnitudes are ($u$, $g$, $r$, $i$, $z$) = (22, 22.2, 22.2, 21.3, 20.5). The data are saturated at about 14 mag in $g$, $r$, and $i$, and about 12 mag in $u$ and $z$ \citep[see][for more detail]{Bilir08}.

2MASS provides the most complete database of near infrared (NIR) Galactic point sources available to date. Observations cover 99.998 per cent \citep{Skrutskie06} of the sky with simultaneous detections in $J$ (1.25 $\mu$m), $H$ (1.65 $\mu$m), and $K_{s}$ (2.17 $\mu$m) bands up to limiting magnitudes of 15.8, 15.1, and 14.3, respectively. Bright source extractions have 1$\sigma$ photometric uncertainty of $<$ 0.03 mag and astrometric accuracy on the order of 100 mas \citep[see][for more detail]{Bilir08}.

Another infrared survey was made by Spitzer Space Telescope which is one of the space baring observatories \citep{Werner04}. Spitzer has various instruments on board and one of them is the Infrared Array Camera (IRAC) which was used for the Galactic Legacy Mid-Plane Survey Extraordinaire (GLIMPSE) programme. The GLIMPSE survey was devised to obtain images of infrared sources toward the inner Galactic plane at 3.6, 4.5, 5.8 and 8.0 $\mu$m with angular resolution between 1$\arcsec$.4 to 1$\arcsec$.9 using IRAC on the Spitzer Space Telescope \citep{Churchwell01,Benjamin05}.

WISE (Wide Field Survey Explorer), the up-to-date infrared survey, began surveying the sky on 14 January 2010 and completed its first full coverage of the sky on 17 July 2010 with much higher sensitivity than comparable previous infrared survey missions \citep{Wright10}. WISE has four infrared filters $W1$, $W2$, $W3$, $W4$ centered at 3.4, 4.6, 12 \& 22 $\mu$m, and the angular resolution is 6$\arcsec$.1, 6$\arcsec$.4, 6$\arcsec$.5, and 12$\arcsec$.0 at $W1$, $W2$, $W3$, $W4$ by using a 40 cm telescope feeding arrays with a total of four million pixels. The increased number of detectors leads to a much higher sensitivity: WISE has achieved five $-\sigma$ point source sensitivities better than 0.08, 0.11, 1 and 6 mJy at 3.4, 4.6, 12 and 22 $\mu$m. These sensitivities correspond to Vega magnitudes 16.5, 15.5, 11.2, and 7.9. Thus WISE will go a magnitude deeper than the 2MASS $K_{s}$ data in $W1$ for sources with spectra close to that of an A0 star, and even deeper for moderately red sources like K stars or galaxies with old stellar populations. WISE filters fill the wavelength gap between 2MASS \citep{Skrutskie06} all sky survey at 1.2-–2.2 $\mu$m and AKARI mission, \citep{Murakami07} which used an ingenious technique to survey the middle IR sky at 9 and 18 $\mu$m with sensitivities of 50 and 100 mJy \citep{Ishihara10} and with better angular resolution than the InfraRed Astronomical Satellite \citep[IRAS;][]{Neugebauer84, Beichman88}. 

The scientific goals of the survey are the Solar System objects (asteroids, comet trails, and zodiac bands), Solar neighbourhood objects (dwarf and giant stars, brown dwarfs, young stars and debris disc, interstellar dust), and the most majority of the extragalactic objects (the most ultra luminous, seyferts, and starburst galaxies, and quasars) in the Universe. The preliminary data (released on 14 April 2011) covers 57 per cent of the sky which are from the first 105 days of WISE survey observations. A source catalogue contains positional and photometric information for over 257 million objects detected on the WISE images, and an Explanatory Supplement that provides a user's guide to the WISE mission and format, content, characteristics and cautionary notes for the release products. 

%FIGURE 1
\begin{figure}
\begin{center}
\includegraphics[scale=0.50, angle=0]{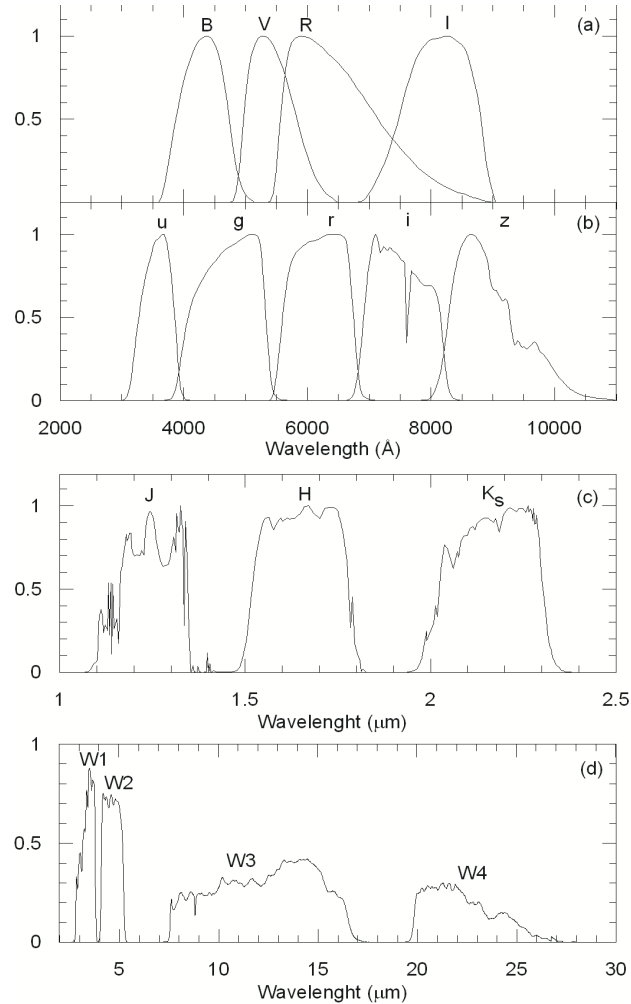}
\caption[] {Normalized passbands of the Johnson-Cousins $BVRI$
filters (a), the {\em SDSS} filters (b), the {\em 2MASS} 
filters (c) and the {\em WISE} filters (d).}
%\label{cmd}
\end{center}
\end{figure}

In \citet{Bilir08} we presented the transformations between {\it BVRI}, SDSS, and 2MASS photometric systems. The passband profiles for $BVRI$, $ugriz$, $JHK_{s}$, and $W1W2W3W4$ photometric systems are given in Fig. 1. Here we add a middle infrared system, WISE, to this photometric set and we follow the same procedure with slight modifications: As the transformations were planned for dwarfs, $W3$ and $W4$ magnitudes were not used due to low energy of dwarfs in those bands. The transformations are metallicity dependent and are a function of two colours, similar to \citet{Bilir08}. However, in the inverse transformations, i.e., from the WISE system to $BVRI$ and $gri$, $J-H$ colour of 2MASS photometric system is used as a second colour combined linearly with $W1-W2$. Thus, ${J-H}$ colour filled the gap of $W3-W4$. Such a modification seems reasonable, as both $J–-H$ and $W3-W4$ colours correspond to the infrared  photometric systems. 

In Section 2 we present the sources of our star sample and the criteria applied to the chosen stars. The transformation equations
are given in Section 3. Finally, in Section 4, we give a summary and conclusion.

\section{Data}
The data used for our transformations were taken from \citet{Bilir08} and the preliminary data release of WISE \citep{Wright10}. \citet{Bilir08} used the original $BVRI$, $gri$, and $JHK_{s}$ magnitudes taken from \citet{Stetson00, Saha05}, \citet{Adelman-McCarthy07}, and \citet{Cutri03}, respectively, and formed a catalogue of 825 dwarfs (the identification of dwarfs are explained in Bilir et al. 2008). We matched the coordinates of these stars in 10 fields (NGC 2419, NGC 2420, Draco, NGC 2683, NGC 3031, L106, L107, Pal 5, PG–1633, M92) investigated by \citet{Stetson00} and mapped by WISE, and revealed that 311 out of 825 stars appeared in the catalogue of \citet{Bilir08} and preliminary data release of WISE\footnote{http://irsa.ipac.caltech.edu/cgi-bin/Gator/nph-scan?mission=irsa\&submit=Select\&projshort=WISE\_PRELIM}. We confirmed the overlapping of 311 stars  by comparing their $J$ magnitudes which appeared in two different sources. We reduced the number of stars to 289 whose magnitudes were of  best quality (AA) for $W1$ and $W2$, and we had to omit all $W3$ and $W4$ magnitudes due to their insufficient quality. A further reduction has been carried out for 72 stars whose $B-V$ and $R-I$ colours were missing in the catalogue of \citet{Bilir08}. The data of our final catalogue for 289 stars are given in Table 1 in electronic format. The positions of the fields (totally 27) investigated by \citet{Stetson00} and those mapped by WISE (10 fields) are given in the equatorial coordinate system in Fig. 2a. Stetson fields with WISE data are also given in the Galactic coordinates in Fig. 2b and the corresponding reduced $E(B-V)$ colour excesses taken from \citet*{Schlegel98} in Fig. 2c.

%Table 1
\begin{table*}
\setlength{\tabcolsep}{2pt}
{\scriptsize
\center
\caption{Johnson-Cousins, SDSS, 2MASS and WISE magnitudes and colours 
of the sample stars (289 total stars). The columns give: (1) Star name;
(2) and (3) Galactic coordinates; (4) $V$--apparent magnitude; (5)
and (6) $(B-V)$ and $(R-I)$ colour indices; (7) $g$--apparent magnitude;
(8) and (9) $(g-r)$, $(r-i)$ colour indices; (10)$J$-apparent magnitude; 
(11) and (12) $(J-H)$, $(H-K_{s})$ colour indices, (13) $W1$-apparent 
magnitude; and (14) $(W1-W2)$ colour index, and (15) reduced $E_{d}(B-V)$ colour 
excess. The complete table is available in electronic format.}
\begin{tabular}{lcccccccccccccc}
\hline
(1) & (2) & (3) & (4) &  (5) &  (6) & (7) &  (8) &  (9) &  (10) &    (11) &
 (12) &  (13) & (14) & (15)\\
Star & $l~(^{\circ})$ & $b~(^{\circ})$ & $V$ &  $(B-V)$ &  $(R-I)$ & $g$ &  
$(g-r)$ &  $(r-i)$ &  $J$ &  $(J-H)$ &  $(H-K_{s})$ & $W1$ & $(W1-W2)$ & $E_{d}(B-V)$ \\
\hline
Pal5-S10 & 0.78766 & 45.85286 & 16.742 & 0.936 &    & 17.209 & 0.765 & 0.337 & 14.782 & 0.447 & 0.059 & 14.162 & -0.023 & 0.058 \\
Pal5-S13 & 0.79079 & 45.85054 & 16.499 & 0.693 &    & 16.824 & 0.553 & 0.241 & 14.941 & 0.442 & -0.218& 14.430 & -0.064 & 0.058 \\
Pal5-S14 & 0.79423 & 45.85242 & 17.798 & 1.120 &    & 18.424 & 1.072 & 0.417 & 15.520 & 0.643 & 0.191 & 14.795 & -0.049 & 0.058 \\
    ...  &    ...   &    ...  &  ...   &   ... & ...&  ...   &   ... &  ...  &  ...   &  ...  &  ...  &   ...  &  ...   & ...   \\           
    ...  &    ...   &    ...  &  ...   &   ... & ...&  ...   &   ... &  ...  &  ...   &  ...  &  ...  &   ...  &  ...   & ...   \\
    ...  &    ...   &    ...  &  ...   &   ... & ...&  ...   &   ... &  ...  &  ...   &  ...  &  ...  &   ...  &  ...   & ...   \\      
L106-S10 & 351.06012& 51.84716& 14.892 & 0.760 &    & 15.227 & 0.561 & 0.219 & 13.444 & 0.517 & 0.041 & 12.886 & -0.012 & 0.039 \\
L106-S13 & 351.06807& 51.81729& 15.321 & 0.618 &    & 15.566 & 0.426 & 0.148 & 14.069 & 0.297 & -0.007& 13.628 & -0.013 & 0.039 \\
L106-S11 & 351.07185& 51.84046& 16.234 & 0.726 &    & 16.550 & 0.548 & 0.208 & 14.771 & 0.429 & 0.117 & 14.237 &  0.035 & 0.039 \\ 
\hline
\end{tabular}}
\end{table*}

%FIGURE 2
\begin{figure}
\begin{center}
\includegraphics[scale=0.50, angle=0]{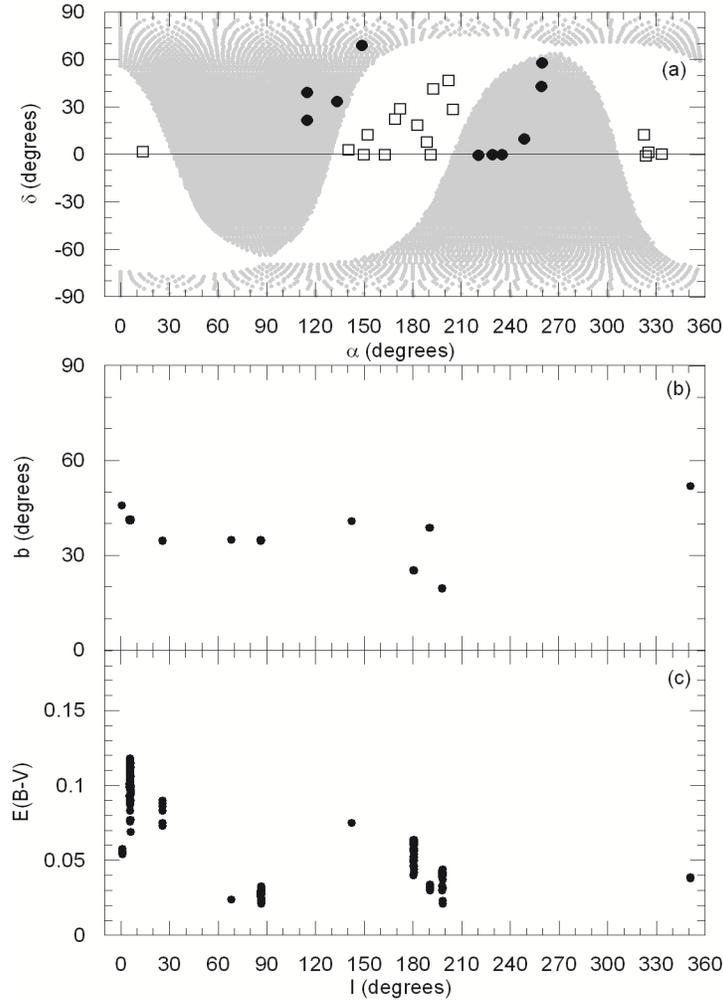}
\caption[] {Equatorial coordinates of Stetson's (2000) fields used by \citet{Bilir08} (a), Galactic coordinates of Stetson's fields with available WISE data (b), and distribution of the colour excess $E(B-V)$, taken from \citet{Schlegel98}, with Galactic longitude (c). Symbols: grey regions correspond to the fields of Stetson with available preliminary WISE data, open squares are the fields without WISE data, and filled circles are the fields with available WISE data.}
\end{center}
\end{figure}

The mean errors, standard deviations, and the number of stars available for magnitudes for four photometric systems are given in Table 2. The mean errors are less than 1 per cent and 2 per cent for the magnitudes $BVRI$ and $gri$, respectively, whereas they lie between 3 per cent and 6 per cent for near and middle infrared photometric systems.  

%Table 2
\begin{table}
\center \caption{Mean errors, standard deviations and number of stars for the filters
of Johnson-Cousins, SDSS, 2MASS and WISE photometries.}
\begin{tabular}{ccccc}
\hline
 Filter & Mean error(mag) & $s$        &  N  & Photometry \\
\hline
         B & 0.0071 &      0.0045 & 246 & {\em BVRI} \\
         V & 0.0042 &      0.0028 & 281 &            \\
         R & 0.0082 &      0.0044 & 263 &            \\
         I & 0.0069 &      0.0039 & 271 &            \\
         g & 0.0145 &      0.0046 & 289 &      SDSS  \\
         r & 0.0136 &      0.0055 & 289 &            \\
         i & 0.0144 &      0.0036 & 289 &            \\
         J & 0.0356 &      0.0106 & 289 &      2MASS \\
         H & 0.0432 &      0.0144 & 289 &            \\
   $K_{s}$ & 0.0591 &      0.0227 & 289 &            \\
        W1 & 0.0322 &      0.0066 & 289 &      WISE  \\
        W2 & 0.0513 &      0.0018 & 289 &            \\

\hline
\end{tabular}
\end{table}

\subsection{Reddening and Metallicity}

Dereddening of the magnitudes is carried out by using the following procedure. First, we used the $R_{\lambda}/R_{3.1}$ data of \citet*{Cardelli89} and obtained a spline function for the infrared wavelengths between 2 and 35 $\mu$m. Then, we used the colour excess $E(B-V)$ values from \citet{Schlegel98} at appropriate Galactic latitudes to evaluated the total extinction $R_{\lambda}$ for the wavelengths 3.4, 4.6, 12, and 22 $\mu$m corresponding to the bands $W1$, $W2$, $W3$, and $W4$, i.e. 0.158, 0.093, 0.087, and 0.056 mag, respectively (Fig. 3). Thus, the magnitudes $W1$ and $W2$ were dereddened by 0.158 and 0.093 mag, respectively. The $BVRI$, $gri$, and $JHK_{s}$ magnitudes had already been dereddened by \citet{Bilir08}. The distribution of errors for 12 magnitudes are given in Fig. 4 as a function of intrinsic colours.

%FIGURE 3
\begin{figure}
\begin{center}
\includegraphics[scale=0.50, angle=0]{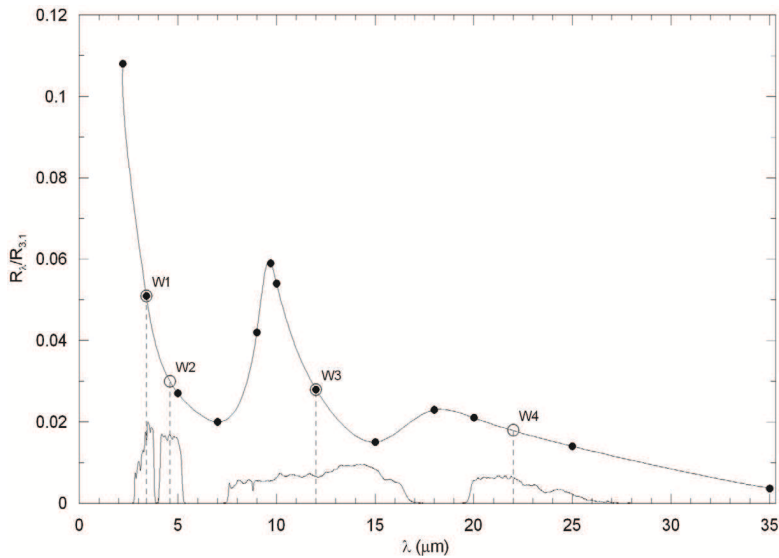}
\caption[] {Typical interstellar extinction curve for the mid infrared normalized to $R_{V}=3.1$, and the passbands of the WISE filters. The filled circles correspond to the values taken from \citet{Cardelli89}, and the solid line represent the spline fit to the data. The dashed lines denote the position of the effective wavelengths of the WISE filters and the open circles correspond to the extinction values of the WISE filters.}
\end{center}
\end{figure}

The transformation formulae are given as a function of metallicity. For this purpose, we used the procedure of \citet*{Karaali05} and determined the metallicities of 289 stars (Table 3). It turned out that 177 stars of our sample are metal rich ($-0.4<[M/H]\leq+0.2$ dex), 43 stars are of intermediate metallicity ($-1.2<[M/H]\leq-0.4$ dex), and 69 stars are metal poor, ($-3<[M/H]\leq-1.2$ dex).

%Table 3
\begin{table}
\center \caption{Metallicity distribution of the sample. Stars with
$(g-r)_{0}>0.95$ mag were assumed to have a metallicity of
$-0.4<[M/H]\leq+0.2$ dex.}
\begin{tabular}{cc}
\hline
Metallicity (dex) & Number of stars \\
\hline
$-0.4<[M/H]\leq+0.2$  & 177 \\
$-1.2<[M/H]\leq -0.4$ &  43 \\
$-3.0<[M/H]\leq -1.2$ &  69 \\
\hline
\end{tabular}
\end{table}

%FIGURE 4
\begin{figure*}
\begin{center}
\includegraphics[scale=0.70, angle=0]{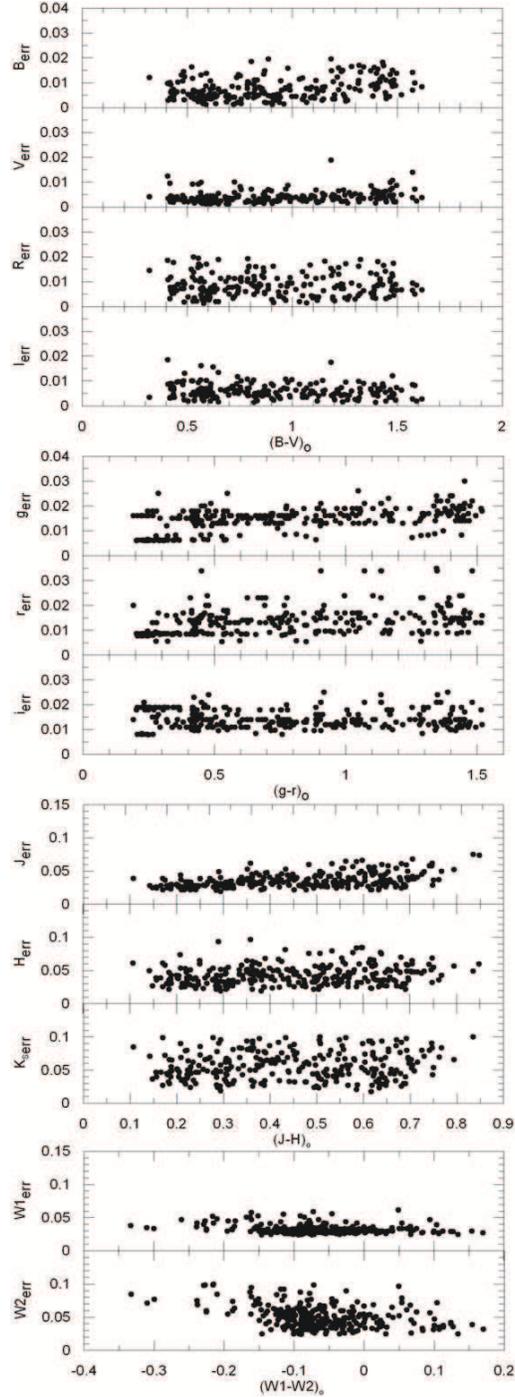}
\caption[] {The error distributions for Johnson-Cousins $BVRI$, SDSS $
gri$ 2MASS $JHK_{s}$ and WISE $W1W2$.}
\end{center}
\end{figure*}

\section{Results}
\subsection{Transformations between WISE and Johnson–-Cousins' Photometry}
We used the following general equations and derived four sets of transformations between the $W1$ and $W2$ magnitudes of WISE and Johnson-–Cousins' $BVRI$. We did not consider the transformations for $W3$ and $W4$ due to the reason explained in Section 2. The transformation sets are as follows: Transformations for the whole sample, metal rich stars, intermediate metallicity stars, and metal poor stars, respectively. The definition of the last three sub samples are already given in the previous section. The general equations are: 

\begin{equation}
(V-–W1)_{0}=a_{1}(B-V)_{0}+b_{1}(B-V)_{0}(R-I)_{0}+c_{1}(R-I)_{0}+d_{1},
\end{equation}

\begin{equation}
(V-–W2)_{0}=a_{2}(B-V)_{0}+b_{2}(B-V)_{0}(R-I)_{0}+c_{2}(R-I)_{0}+d_{2}.
\end{equation}

The transformation equations consist of two colours but three terms. The term $b_i$ $(B-V)_0$ $(R-I)_0$ (i= 1, 2) provides more accurate colours and results residuals less than the ones obtained considering only two terms. The numerical values of the coefficients $a_i$, $b_i$, $c_i$, and $d_i$ (i=1, 2) for the four sets are given in columns (1) and (2) in Table 4. The fifth and sixth numbers in each column are the squared correlation coefficient and the standard deviation for the colour indicated at the top of the column. There are similarities between the values of the coefficients evaluated for different samples, except the ones for intermediate metallicity stars. The metallicity distribution of these sub samples ($-0.4<[M/H]\leq+0.2$, $-1.2<[M/H]\leq-0.4$, and $-3.0<[M/H]\leq-1.2$ dex) resemble the metallicities ranges for the Galactic thin and thick discs or the halo, respectively. Thus the transformations are luminosity and metallicity dependent.  

%Table 4
\begin{table*}
\setlength{\tabcolsep}{3.5pt}
\center \caption{Coefficients $a_{i}$, $b_{i}$, $c_{i}$ and $d_{i}$ for the
transformation Eqs. 1--6, in column matrix form
for the four different metallicity ranges. The subscript $i$=1, 2, 3, 4, 5 
and 6 correspond to the same number that denotes the columns. Numerical
values in the fifth, sixth, and seventh lines of each metallicity range are 
the squared correlation coefficients ($R^{2}$), the standard deviations 
($s$) and number of stars (N), respectively.}

\begin{tabular}{cccccccc}
\hline
           &       &        (1) &        (2)&        (3) &        (4)&        (5) &        (6) \\
\hline
$[M/H]$ (dex)&            & $(V-W1)_{o}$ & $(V-W2)_{o}$ & $(g-W1)_{o}$ & $(g-W2)_{o}$ & $(J-W1)_{o}$ & $(J-W2)_{o}$\\
\hline
[-3,+0.2)  &     $a_{i}$ & 1.754 $\pm$ 0.048 &  1.596 $\pm$ 0.056 & 1.987 $\pm$ 0.031 &  1.916 $\pm$ 0.037 & 1.246 $\pm$ 0.041 &  1.225 $\pm$ 0.064 \\
           &     $b_{i}$ &-0.665 $\pm$ 0.116 & -0.395 $\pm$ 0.136 &-1.161 $\pm$ 0.107 & -0.971 $\pm$ 0.127 & 0.399 $\pm$ 0.313 &  1.458 $\pm$ 0.491 \\
           &     $c_{i}$ & 2.742 $\pm$ 0.200 &  2.549 $\pm$ 0.236 & 3.487 $\pm$ 0.176 &  3.418 $\pm$ 0.207 & 0.377 $\pm$ 0.159 &  0.025 $\pm$ 0.249 \\
           &     $d_{i}$ &-0.214 $\pm$ 0.061 & -0.189 $\pm$ 0.072 & 0.579 $\pm$ 0.022 &  0.519 $\pm$ 0.026 & 0.127 $\pm$ 0.018 &  0.052 $\pm$ 0.029 \\
           &     $R^{2}$ & 0.991 & 0.989 & 0.994 & 0.992 & 0.906 & 0.823\\
           &         $s$ & 0.096 & 0.113 & 0.100 & 0.118 & 0.079 & 0.125\\
           &          N  & 217 & 217  & 289 & 289 & 289 & 289 \\
\hline
[-0.4,+0.2)&     $a_{i}$& 1.737 $\pm$ 0.061 &  1.565 $\pm$ 0.070 & 1.975 $\pm$ 0.034 &  1.903 $\pm$ 0.040 & 1.345 $\pm$ 0.048 &  1.358 $\pm$ 0.084 \\
           &     $b_{i}$&-0.669 $\pm$ 0.167 & -0.231 $\pm$ 0.192 &-1.136 $\pm$ 0.138 & -0.873 $\pm$ 0.162 &-0.398 $\pm$ 0.407 &  0.356 $\pm$ 0.709 \\
           &     $c_{i}$& 2.782 $\pm$ 0.287 &  2.305 $\pm$ 0.330 & 3.454 $\pm$ 0.220 &  3.283 $\pm$ 0.258 & 0.878 $\pm$ 0.224 &  0.720 $\pm$ 0.390 \\
           &     $d_{i}$&-0.230 $\pm$ 0.085 & -0.139 $\pm$ 0.098 & 0.594 $\pm$ 0.027 &  0.534 $\pm$ 0.031 & 0.073 $\pm$ 0.023 & -0.020 $\pm$ 0.040 \\
           &     $R^{2}$& 0.992 & 0.990 & 0.996 & 0.994 & 0.928 & 0.838 \\
           &        $s$ & 0.102 & 0.118 & 0.100 & 0.118 & 0.078 & 0.137 \\
           &          N & 120 & 120 & 177 & 177 & 177 & 177 \\
\hline
[-1.2,-0.4)&     $a_{i}$ &  2.443 $\pm$ 0.502 &  2.225 $\pm$ 0.600 &  2.157 $\pm$ 0.274 &  2.316 $\pm$ 0.329 &  0.735 $\pm$ 0.198 &  0.474 $\pm$ 0.227 \\
           &     $b_{i}$ & -1.768 $\pm$ 1.086 & -1.482 $\pm$ 1.296 & -2.087 $\pm$ 1.022 & -2.370 $\pm$ 1.229 &  2.242 $\pm$ 1.705 &  4.629 $\pm$ 1.953 \\
           &     $c_{i}$ &  2.731 $\pm$ 0.731 &  2.651 $\pm$ 0.873 &  3.669 $\pm$ 0.614 &  3.490 $\pm$ 0.738 & -0.371 $\pm$ 0.542 & -1.153 $\pm$ 0.620 \\
           &     $d_{i}$ & -0.383 $\pm$ 0.287 & -0.366 $\pm$ 0.342 &  0.512 $\pm$ 0.110 &  0.408 $\pm$ 0.133 &  0.301 $\pm$ 0.069 &  0.308 $\pm$ 0.079 \\
           &    $R^{2}$  & 0.958 & 0.939 & 0.969 & 0.954 & 0.634 &  0.572\\
           &        $s$  & 0.075 & 0.089 & 0.079 & 0.095 & 0.071 &  0.081\\
           &          N &  37 & 37 & 43& 43 & 43 & 43 \\
\hline
[-3,-1.2)  &    $a_{i}$ &  1.567 $\pm$ 0.329 &  1.568 $\pm$ 0.410 & 1.847  $\pm$ 0.223 &  1.828 $\pm$ 0.262  & 1.084 $\pm$ 0.072 &  1.019 $\pm$ 0.103 \\
           &    $b_{i}$ &  0.078 $\pm$ 0.669 & -0.086 $\pm$ 0.835 & -0.841 $\pm$ 0.823 & -1.080 $\pm$ 0.967 & -0.222 $\pm$ 0.730 &  1.051 $\pm$ 1.046\\
           &    $c_{i}$ &  1.865 $\pm$ 0.628 &  2,166 $\pm$ 0.783 &  3.451 $\pm$ 0.750 &  3.782 $\pm$ 0.881 &  0.380 $\pm$ 0.331 & -0.165 $\pm$ 0.474\\
           &    $d_{i}$ &  0.090 $\pm$ 0.252 & -0.065 $\pm$ 0.314 &  0.639 $\pm$ 0.128 &  0.538 $\pm$ 0.150 &  0.220 $\pm$ 0.033 &  0.169 $\pm$ 0.048\\
           &    $R^{2}$ &  0.972 &      0.957 &      0.971 &      0.961 &      0.861 &      0.759\\
           &        $s$ &  0.087 &      0.109 &      0.107 &      0.126 &      0.061 &      0.087 \\
           &          N & 60 & 60 & 69 & 69 & 69 & 69 \\
\hline
\end{tabular}
\end{table*}

\subsection{Transformations between WISE and SDSS photometry}
The transformations between the $W1$ and $W2$ magnitudes of WISE and SDSS have similar general equations, given below:

\begin{equation}
(g-–W1)_{0}=a_{3}(g-r)_{0}+b_{3}(g-r)_{0}(r-i)_{0}+c_{3}(r-i)_{0}+d_{3},
\end{equation}

\begin{equation}
(g-–W2)_{0}=a_{4}(g-r)_{0}+b_{4}(g-r)_{0}(r-i)_{0}+c_{4}(r-i)_{0}+d_{4}.
\end{equation}

The numerical values of the coefficients $a_i$, $b_i$, $c_i$ and $d_i$ (i=3, 4) for the four sets defined above are given in columns (3) and (4) in Table 4. Again, the transformations between the magnitudes $W1$ and $W2$ of WISE and SDSS photometry are luminosity, metallicity and two colour dependent.

\subsection{Transformations between WISE and 2MASS photometry} 
The transformations between the $W1$ and $W2$ magnitudes of WISE and 2MASS have similar general equations, given below:

\begin{equation}
(J-W1)_{0}=a_{5}(J-H)_{0}+b_{5}(J-H)_{0}(H-K_{S})_{0}+c_{5}(H-K_{S})_{0}+d_{5},
\end{equation}

\begin{equation}
(J-W2)_{0}=a_{6}(J-H)_{0}+b_{6}(J-H)_{0}(H-K_{S})_{0}+c_{6}(H-K_{S})_{0}+d_{6}.
\end{equation}

The numerical values of the coefficients $a_i$, $b_i$, $c_i$ and $d_i$ (i=5, 6) for the four sets defined above are given in columns (5) and (6) in Table 4. As it was obtained both for Johnson–-Cousins' and SDSS photometry, the transformations between the magnitudes $W1$ and $W2$ of WISE and 2MASS are luminosity, metallicity and two colour dependent.

\subsection {Residuals}
We compared the observed colours and those evaluated via Eqs. (1)-(6) to analyze the residuals distribution. As seen in Table 5, the mean of the residuals are rather small. The residuals are plotted versus observed $(B-V)_{0}$, $(g-r)_{0}$ or $(J-H)_{0}$ colours in Fig. 5. Although the number of stars are not equal in each panel, there is no systematic deviation from the zero point in any panel. However, the ranges of the residuals for different colours are not the same. Those for $V-W1$, $g-W1$, and $J-W1$ are smaller than the ones for $V-W2$, $g-W2$, and $J-W2$. This indicates that the $W1$ (absolute) magnitudes evaluated via the transformations given above would be more accurate than $W2$.

%Table 5
\begin{table*}
\setlength{\tabcolsep}{2pt}
\center
\caption{Averages for differences between the measured and evaluated 
colours (residuals) for six colours in four different metallicity ranges. 
The notation used is $\Delta$(colour) = (evaluated colour) - 
(measured colour).}
\begin{tabular}{ccccccc}
\hline
$[M/H]$ (dex) & $<\Delta (V-W1)_o>$ & $<\Delta (V-W2)_o>$ & $<\Delta (g-W1)_o>$ & $<\Delta (g-W2)_o>$ & $<\Delta (J-W1)_o>$ & $<\Delta (J-W2)_o>$ \\
\hline
$[-3.0,+0.2)$   & -0.00006 &     0.00003 &    0.00000 &   -0.00002 &     0.00061 &      0.00191 \\
$[-0.4,+0.2)$   & -0.00008 &    -0.00006 &    0.00002 &   -0.00002 &    -0.00001 &     -0.00001 \\
$[-1.2,-0.4)$   & -0.00000 &     0.00000 &    0.00012 &   -0.00002 &     0.00002 &      0.00002 \\
$[-3.0,-1.2)$   &  0.00000 &     0.00000 &   -0.00007 &    0.00000 &     0.00001 &     -0.00007 \\
\hline
\end{tabular}
\end{table*}

%FIGURE 5
\begin{figure}
\begin{center}
\includegraphics[scale=0.80, angle=0]{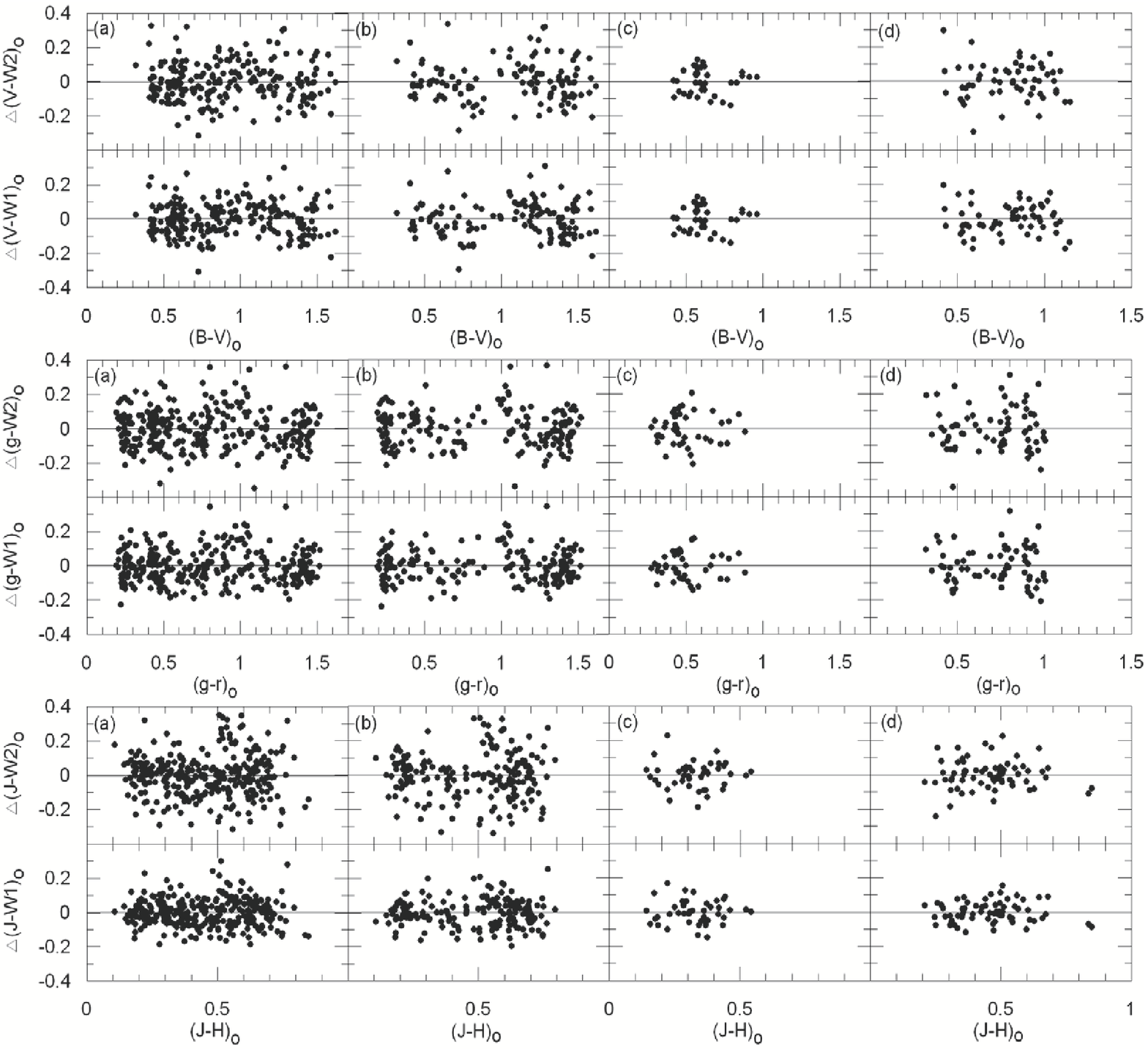}
\caption[] {Colour residuals for different metallicity ranges. (a) for the whole sample $-3<[M/H]\leq+0.2$, (b) for high metallicity $-0.4<[M/H]\leq+0.2$, (c) for intermediate metallicity $-1.2<[M/H]\leq-0.4$, and (d) for low metallicity $-3<[M/H]\leq-1.2$ dex. The notation used is $\Delta$(colour)=(evaluated colour)-(adopted colour).}
\end{center}
\end{figure}

\subsection {Inverse transformation formulae}
As it was explained before, $W3$ and $W4$ magnitudes can not be used for the star sample. Hence, we adapted the following procedure to get the inverse transformations with two colours: By combining linearly the near and mid infrared colours, i.e. $(J-H)_{0}$ and $(W1-W2)_{0}$, we transformed them to the optical colours, i.e. $(B-V)_{0}$, $(R-I)_{0}$, $(g-r)_{0}$, and $(r-i)_{0}$. Additionally in Table 6, we separated the stars in the four metallicity ranges defined above into bins and used the locus of each bin for numerical evaluation of the coefficients in the inverse transformations given in the following: 

\begin{equation}
(B-V)_{0}=\alpha_{1}(J-H)_{0}+\beta_{1}(W1-W2)_{0}+\gamma_{1},
\end{equation}

\begin{equation}
(R-I)_{0}=\alpha_{2}(J-H)_{0}+\beta_{2}(W1-W2)_{0}+\gamma_{2},
\end{equation}

\begin{equation}
(g-r)_{0}=\alpha_{3}(J-H)_{0}+\beta_{3}(W1-W2)_{0}+\gamma_{3},
\end{equation}

\begin{equation}
(r-i)_{0}=\alpha_{4}(J-H)_{0}+\beta_{4}(W1-W2)_{0}+\gamma_{4}.
\end{equation}

The numerical values of the coefficients, $\alpha_i$, $\beta_i$, and $\gamma_i$ (i=1, 2, 3, 4) for the four metallicity ranges are given in Table 7.

%Table 6
\begin{table*}
\setlength{\tabcolsep}{2pt}
\scriptsize{
\center
\caption{$(B-V)_{o}$ and $(g-r)_{o}$ bins used for the inverse transformation equations for Johnson - Cousins BVRI and SDSS gri, for the metallicity ranges $-3<[M/H]\leq+0.2$, $-0.4<[M/H]\leq+0.2$, $-1.2<[M/H]\leq-0.4$, and $-3<[M/H]\leq-1.2$ dex. $N$ is the number of stars in each bin. The other columns refer to the colours used in the transformations.} 
\begin{tabular}{cccccccccccccc}
\hline
$[M/H]$ (dex)&      $(B-V)_{o}$ &          N &   $(B-V)_{o}$ &   $(R-I)_{o}$ &  $(J-H)_{o}$ &   $(W1-W2)_{o}$& $[M/H]$ (dex)& $(g-r)_{o}$ & N &  $(g-r)_{o}$ & $(r-i)_{o}$ & $(J-H)_{o}$ & $(W1-W2)_{o}$ \\
\hline
$[-3.0,+0.2)$	& [0.3,0.5] &      22 &      0.434 &      0.284 &      0.214 &     -0.056 & $[-3.0,+0.2)$ & [0.1,0.3] &      37 &      0.245 &      0.079 &      0.216 &     -0.071 \\
		& (0.5,0.7] &      57 &      0.581 &      0.376 &      0.321 &     -0.084 &            & (0.3,0.5] &         65 &      0.424 &      0.144 &      0.294 &     -0.084 \\
		& (0.7,0.9] &      39 &      0.816 &      0.476 &      0.437 &     -0.090 &            & (0.5,0.7] &         36 &      0.583 &      0.205 &      0.398 &     -0.090 \\
		& (0.9,1.1] &      30 &      0.988 &      0.568 &      0.544 &     -0.087 &            & (0.7,0.9] &         43 &      0.777 &      0.287 &      0.501 &     -0.087 \\
		& (1.1,1.3] &      29 &      1.202 &      0.673 &      0.639 &     -0.082 &            & (0.9,1.1] &         32 &      0.996 &      0.404 &      0.575 &     -0.091 \\
		& (1.3,1.5] &      34 &      1.413 &      1.003 &      0.640 &     -0.025 &            & (1.1,1.3] &         28 &      1.185 &      0.497 &      0.639 &     -0.082 \\
		& (1.5,1.7] &       6 &      1.579 &      1.249 &      0.562 &      0.098 &            & (1.3,1.5] &         46 &      1.404 &      0.902 &      0.624 &      0.028 \\
$[-0.4,+0.2)$  	& [0.3,0.5] &      13 &      0.435 &      0.261 &      0.216 &     -0.053 &            & (1.5,1.7] &          2 &      1.517 &      1.239 &      0.552 &      0.133 \\
		& (0.5,0.7] &      17 &      0.609 &      0.382 &      0.294 &     -0.090 & $[-0.4,+0.2)$ & [0.1,0.3] &      35 &      0.242 &      0.079 &      0.216 &     -0.072 \\
		& (0.7,0.9] &      15 &      0.788 &      0.440 &      0.402 &     -0.096 & 	       & (0.3,0.5] &         21 &      0.413 &      0.144 &      0.284 &     -0.090 \\
		& (0.9,1.1] &       8 &      1.054 &      0.558 &      0.561 &     -0.082 &            & (0.5,0.7] &         17 &      0.591 &      0.210 &      0.387 &     -0.091 \\
		& (1.1,1.3] &      27 &      1.215 &      0.676 &      0.639 &     -0.084 &            & (0.7,0.9] &         12 &      0.758 &      0.281 &      0.501 &     -0.106 \\
		& (1.3,1.5] &      34 &      1.413 &      1.003 &      0.640 &     -0.014 &            & (0.9,1.1] &         16 &      1.045 &      0.424 &      0.623 &     -0.099 \\
		& (1.5,1.7] &       6 &      1.579 &      1.249 &      0.562 &      0.098 &            & (1.1,1.3] &         28 &      1.185 &      0.497 &      0.639 &     -0.082 \\
$[-1.2,-0.4)$ 	& [0.3,0.5] &       6 &      0.447 &      0.292 &      0.180 &     -0.060 &            & (1.3,1.5] &         46 &      1.404 &      0.902 &      0.624 &      0.028 \\
		& (0.5,0.7] &      23 &      0.571 &      0.362 &      0.341 &     -0.084 &            & (1.5,1.7] &          2 &      1.517 &      1.239 &      0.552 &      0.134 \\
		& (0.7,0.9] &       6 &      0.809 &      0.462 &      0.412 &     -0.096 & $[-1.2,-0.4)$ & [0.1,0.3] &       2 &      0.283 &      0.082 &      0.205 &     -0.052 \\
		& (0.9,1.1] &       2 &      0.935 &      0.527 &      0.533 &     -0.084 &            & (0.3,0.5] &         26 &      0.429 &      0.150 &      0.304 &     -0.080 \\
$[-3.0,-1.2)$	& [0.3,0.5] &       3 &      0.427 &      0.376 &      0.271 &     -0.066 &            & (0.5,0.7] &          9 &      0.549 &      0.199 &      0.372 &     -0.107 \\
		& (0.5,0.7] &      17 &      0.566 &      0.381 &      0.332 &     -0.061 &            & (0.7,0.9] &          6 &      0.773 &      0.274 &      0.453 &     -0.088 \\
		& (0.7,0.9] &      18 &      0.821 &      0.498 &      0.474 &     -0.079 & $[-3.0,-1.2)$ & (0.3,0.5] &      18 &      0.445 &      0.140 &      0.308 &     -0.080 \\
		& (0.9,1.1] &      20 &      0.977 &      0.574 &      0.543 &     -0.088 &            & (0.5,0.7] &         10 &      0.595 &      0.212 &      0.392 &     -0.078 \\
		& (1.1,1.3] &       2 &      1.132 &      0.625 &      0.602 &     -0.044 &            & (0.7,0.9] &         25 &      0.784 &      0.301 &      0.505 &     -0.078 \\
		&           &         &            &            &            &            &            & (0.9,1.1] &         16 &      0.947 &      0.365 &      0.530 &     -0.088 \\
\hline
\end{tabular}
}  
\end{table*}

%Table 7
\begin{table*}
\setlength{\tabcolsep}{3.5pt}
\center
\caption{Numerical values of the coefficients for the inverse transformation equations for the different metallicity ranges. $\alpha_{i}$, $\beta_{i}$ and $\gamma_{i}$ ($i$=1, 2, 3 and 4) correspond to equations 7, 8, 9, and 10, respectively. The numerical values in the fourth and fifth lines are the squared correlation coefficients ($R^{2}$) and the standard deviations ($s$). $N$ is the number of stars used in the evaluation of the coefficients.}
\begin{tabular}{cccccc}
\hline
           &       &        (1) &        (2)&		(3)&		(4) \\
\hline
$[M/H]$ (dex)&            & $(B-V)_{o}$ & $(R-I)_{o}$ & $(g-r)_{o}$ & $(r-i)_{o}$\\
\hline
[-3.0,+0.2)&     $\alpha_{i}$ & 2.002  $\pm$ 0.064 &  1.198 $\pm$ 0.142 &  2.203 $\pm$ 0.103 & 1.050  $\pm$ 0.065 \\
           &     $\beta_{i}$  & 2.898  $\pm$ 0.155 &  3.514 $\pm$ 0.340 &  2.614 $\pm$ 0.399 & 3.882  $\pm$ 0.252 \\
           &     $\gamma_{i}$ & 0.177  $\pm$ 0.035 &  0.250 $\pm$ 0.077 & -0.042 $\pm$ 0.064 & 0.137  $\pm$ 0.041 \\
           &     $R^{2}$      & 0.998  &   0.984   &  0.994 & 0.995 \\
           &         $s$      & 0.025  &   0.055   &  0.039 & 0.025 \\
           &          N       &   217  &    217    &   289  &   289 \\
	 
\hline
[-0.4,+0.2)&     $\alpha_{i}$ & 1.914 $\pm$ 0.132 &  1.122 $\pm$ 0.191 &  2.137 $\pm$ 0.083 & 1.028  $\pm$ 0.051\\
           &     $\beta_{i}$  & 2.632 $\pm$ 0.323 &  3.461 $\pm$ 0.469 &  2.588 $\pm$ 0.160 & 3.783  $\pm$ 0.098\\
           &     $\gamma_{i}$ & 0.228 $\pm$ 0.071 &  0.280 $\pm$ 0.103 & -0.005 $\pm$ 0.044 & 0.159  $\pm$ 0.027\\
           &     $R^{2}$      & 0.990 &   0.970   &  0.996 & 0.998 \\
           &        $s$       & 0.052 &   0.076   &  0.034 & 0.021 \\
           &         N        &  120  &    120    &   177  &  177  \\
\hline
[-1.2,-0.4)&     $\alpha_{i}$ & 1.454 $\pm$ 0.603 &  0.693 $\pm$ 0.227 &  2.315 $\pm$ 0.156 & 0.838  $\pm$ 0.023\\
           &     $\beta_{i}$  & 0.050 $\pm$ 5.587 &  0.019 $\pm$ 2.220 &  2.190 $\pm$ 0.718 & 0.416  $\pm$ 0.108\\
           &     $\gamma_{i}$ & 0.162 $\pm$ 0.347 &  0.158 $\pm$ 0.131 & -0.085 $\pm$ 0.039 &-0.069  $\pm$ 0.006\\
           &     $R^{2}$      & 0.930 &  0.955    &  0.998 & 0.999 \\
           &         $s$      & 0.102 &  0.038    &  0.018 & 0.003 \\
           &          N       &  37   &    37     &  43    &   43  \\
\hline
[-3.0,-1.2) &   $\alpha_{i}$  & 2.061  $\pm$ 0.045 & 0.810  $\pm$ 0.077 &  1.832 $\pm$ 0.095 &  0.857  $\pm$ 0.041\\
           &    $\beta_{i}$   & 0.975  $\pm$ 0.368 & 0.157  $\pm$ 0.633 & -10.952$\pm$ 2.047 & -3.928  $\pm$ 0.892\\
           &    $\gamma_{i}$  &-0.065  $\pm$ 0.033 & 0.139  $\pm$ 0.057 & -0.989 $\pm$ 0.151 & -0.435  $\pm$ 0.066\\
           &    $R^{2}$       &  0.999 &  0.982 & 0.998 & 0.999 \\
           &        $s$       &  0.012 &  0.021 & 0.015 & 0.006 \\
           &         N        &   60   &  60    &  69   &   69  \\
\hline
\end{tabular}
\end{table*}

\section {Summary and Conclusion}

We presented the colour transformations for the conversion of the $W1$ and $W2$ magnitudes of WISE into three photometric systems, i.e. Johnson-Cousins' $BVRI$, SDSS $gri$, and 2MASS for dwarfs. $W3$ and $W4$ magnitudes were not included due to their insufficient photometric quality (however, transformations to convert of all magnitudes into the three photometric systems used here will be a second subject for giants in a forthcoming paper). The following constraints were applied to the sample taken from \citet{Bilir08}, to obtain the most accurate transformations: 1) the data were dereddened, 2) giants have been identified and excluded from the sample, 3) sample stars have been selected according to data quality, 4) transformations have been derived for sub samples of different metallicity ranges, and 5) transformations are two colour dependent.

The squared correlation coefficients ($R^2$) for the transformations carried out for the four categories (the whole sample, the metal rich, the intermediate metallicity, and the metal poor stars) are rather high with three exceptions. $R^2$ is equal to 0.634 and 0.572 in the transformations of $(J-W1)_0$ and $(J-W2)_0$ for the intermediate metallicity stars, respectively, while is as small as 0.759 in the $(J-W2)_0$ transformation for the metal poor stars. 

The transformation equations have been designed with two colours but with three terms (plus the constant). That is, each transformation equation consists of the linear combination of two colours plus a quadratic term corresponding to their multiplication. The coefficients of the linear colour terms in most transformation equations are compatible with each other, indicating that transformations are two colour dependent indeed. The coefficient of the quadratic term in the Eqs. (1)-(6) is smaller than the ones in the linear terms. However, it provides a better $R^2$ and smaller residuals.

There are similarities between the corresponding coefficients for metal rich and metal poor stars, whereas the coefficients for intermediate metallicity stars are different for other categories. Hence, we can conclude that transformations are metallicity dependent. Additionally, it is also remarkable that the metallicity ranges for three sub samples: $-0.4<[M/H]\leq+0.2$, $-1.2<[M/H]\leq-0.4$, and $-3< [M/H]\leq-1.2$ dex, correspond to the metallicity values assumed for the thin disc, thick disc, and halo, respectively. Hence, the transformations are also luminosity dependent.                  

{\bf Conclusion:} WISE will go a magnitude deeper than the 2MASS $K_{s}$ data in $W1$ for sources with spectra close to that of an A0 star, and even deeper for K and M spectral type stars. 
The present transformations can be applied to stars with known absolute $V$, $g$, or $J$ magnitudes and absolute magnitudes for $W1$ can be provided. We can use these two advantages to investigate the Galactic (thin and thick) discs more accurately. As for example, for finding the answers of some of the following questions: Do the late type dwarfs, such as brown dwarfs, obey the two exponential space density law for discs that we use today? If so, will the Galactic model parameters that would be estimated with the addition of the brown dwarfs and the ones appearing in the literature be similar? The answers of such questions will be helpful in our better understanding of the Galactic discs structure. Moreover, the different metallicity ranges correspond to different Galactic populations, making the relationships a powerful tool in disentangling the Galactic structure. They will be rendered even more powerful when a similar relation is obtained for giant stars (Paper II, in preparation).

\section{Acknowledgments}
We would like to thank the referee Dr. Jay Holberg for
his useful comments that improved the readability of this paper.

This research used the facilities of the Canadian Astronomy Data Centre 
operated by the National Research Council of Canada with the support of 
the Canadian Space Agency.

This research has made use of the NASA/IPAC Infrared Science Archive, 
which is operated by the Jet Propulsion Laboratory, California 
Institute of Technology, under contract with the National Aeronautics 
and Space Administration.

This publication makes use of data products from the Two Micron All
Sky Survey, which is a joint project of the University of
Massachusetts and the Infrared Processing and Analysis
Center/California Institute of Technology, funded by the National
Aeronautics and Space Administration and the National Science
Foundation.

This research has made use of the SIMBAD, and NASA\rq s Astrophysics 
Data System Bibliographic Services.

S. Karaali is grateful to the Beykent University for financial support.


\begin{thebibliography}{99}

\bibitem[Adelman-McCarthy et al.(2007)]{Adelman-McCarthy07}
Adelman-McCarthy J. K., et al., 2007, VizieR On-line Data Catalog:
II/276

\bibitem[Aihara et al.(2011)]{Aihara11}
Aihara H., et al., 2011, ApJS, 193, 29

\bibitem[Beichman et al.(1988)]{Beichman88}
Beichman C. A., Neugebauer G., Habing H. J., Clegg P. E., Chester T. J., 1988, 
Infrared astronomical satellite (IRAS) catalogs and atlases. Volume 1: Explanatory supplement

\bibitem[Bilir et al.(2008)]{Bilir08}
Bilir S., Ak S., Karaali S., Cabrera-Lavers A., Chonis T. S., Gaskell C. M., 
2008, MNRAS, 384, 1178 

\bibitem[Benjamin et al.(2005)]{Benjamin05}
Benjamin R. A., et al., 2005, ApJ, 630L, 149

\bibitem[Cardelli et al.(1989)Cardelli, Clayton \& Mathis]{Cardelli89}
Cardelli J. A., Clayton G. C., Mathis J. S., 1989, ApJ, 345, 245

\bibitem[Churchwell et al.(2001)]{Churchwell01}
Churchwell E., et al., American Astronomical Society, 198th AAS Meeting, \#25.04; 
Bulletin of the American Astronomical Society, 33, 821

\bibitem[Cutri et al.(2003)]{Cutri03}
Cutri R. M., et al., 2003, 2MASS All-Sky Catalog of Point Sources, 
CDS/ADC Electronic Catalogues, 2246

\bibitem[Fukugita et al.(1996)]{Fukugita96}
Fukugita M., Ichikawa T., Gunn J. E., Doi M., Shimasaku K.,
Schneider D. P., 1996, AJ, 111, 1748

\bibitem[Hogg et al.(2001)]{Hogg01}
Hogg D. W., Finkbeiner D. P., Schlegel D. J., Gunn J. E., 2001,
AJ, 122, 2129

\bibitem[Gunn et al.(1998)]{Gunn98}
Gunn J. E., et al., 1998, AJ, 116, 3040

\bibitem[Ishihara et al.(2010)]{Ishihara10}
Ishihara D., et al., 2010, A\&A, 514A, 1

\bibitem[Karaali et al.(2005)Karaali, Bilir \& Tun\c{c}el]{Karaali05}
Karaali S., Bilir S., Tun\c{c}el S., 2005, PASA, 22, 24

\bibitem[Murakami et al.(2007)]{Murakami07}
Murakami H., et al., 2007, PASJ, 59, 369

\bibitem[Neugebauer et al.(1984)]{Neugebauer84}
Neugebauer G., et al., 1984, ApJ, 278L, 1 

\bibitem[Saha et al.(2005)]{Saha05}
Saha A., Dolphin A. E., Thim F., Whitmore B., 2005, PASP, 117, 37

\bibitem[Schlegel et al.(1998)Schlegel, Finkbeiner \& Davis]{Schlegel98}
Schlegel D. J., Finkbeiner D. P., Davis M., 1998, ApJ, 500, 525

\bibitem[Skrutskie et al.(2006)]{Skrutskie06}
Skrutskie M. F., et al., 2006, AJ, 131, 1163

\bibitem[Smith et al.(2002)]{Smith02}
Smith J. A., et al., 2002, AJ, 123, 2121

\bibitem[Steinmetz et al.(2006)]{Steinmetz06}
Steinmetz M., et al., 2006, AJ, 132, 1645

\bibitem[Stetson(2000)]{Stetson00}
Stetson P. B., 2000, PASP, 112, 925

\bibitem[Werner et al. (2004)]{Werner04}
Werner M. W., et al., 2004, ApJS, 154, 1

\bibitem[Wright et al. (2010)]{Wright10}
Wright E. L., et al., 2010, AJ, 140, 1868

\bibitem[York et al. (2000)]{York00}
York D. G., et al., 2000, AJ, 120, 1579

\end{thebibliography}
\end{document}